\documentclass[letter,twocolumn]{revtex4}
\usepackage{bm,graphicx}

\begin{document}

\title{Asymmetric exchange between electron spins in coupled semiconductor quantum dots}
\author{ \c{S}.~C. B\u{a}descu$^1$, Y. B. Lyanda-Geller$^{1,2}$, T. L. Reinecke$^1$\\
\it{$^1$Naval Research Laboratory,  Washington DC 20375\\
$^2$Department of Physics, Purdue University, West Lafayette, IN
47907}}
\date{\today}

\begin{abstract}
We obtain a microscopic description of the interaction between
electron spins in bulk semiconductors and in pairs of
semiconductor quantum dots. Treating the ${\bf k}$$\cdot$$\hat{\bf
p}$ band mixing and the Coulomb interaction on the same footing,
we obtain in the third order an asymmetric contribution to the
exchange interaction arising from the coupling between the spin of
one electron and the relative orbital motion of the other. This
contribution does not depend on the inversion asymmetry of the
crystal and does not conserve the total spin. We find that this
contribution is $\sim$$10^{-3}$ of the isotropic exchange, and is
of interest in quantum information. Detailed evaluations of the
asymmetric exchange are given for several quantum dot systems.
\end{abstract}

 \keywords{} \maketitle

Spin in quantum dots (QDs) are attractive candidates for qubits in
quantum information in part because of their long coherence times
\cite{Gupta99}. Controllable coupling between these spins is an
essential requirement for two qubit quantum gates, and it has been
the focus of much recent research \cite{Loss98}-\cite{Wu02}.
Typically the spin-spin coupling is dominated by the isotropic
exchange interaction $J\,{\bf S}_1$$\cdot$${\bf S}_2$, which
arises from the Coulomb interaction and the Pauli principle. The
isotropy of this part of the exchange implies that the total spin
is conserved. This is important in gate operations, which involve
pulsing $J(t)$ \cite{Stepanenko}.

In general, spin-orbit coupling in solids also gives rise to terms
that are asymmetric between the spins, and which do not conserve
total spin. These terms can cause loss of fidelity in gate
operations involving two spins. To date, the asymmetric terms that
have been discussed involve spin-orbit coupling of individual
electrons in the conduction band \cite{Kavokin01,Gorkov2003}. For
systems with bulk inversion asymmetry, this coupling has the
Dresselhaus form \cite{Dresselhaus,Altshuler}. In addition,
heterostructure asymmetry can introduce the so-called Rashba
coupling \cite{Rashba84}. These two couplings can give an
additional interaction between the spins of the
Dzyaloshinskii-Moryia (DM) form ${\bm \beta}_{so}$$\cdot$$(\,{\bf
S}_1$$\times$${\bf S}_2)$$/\sqrt{\beta_{so}^2+J^2}$ \cite{DM1,DM2}
where ${\bm \beta}_{so}$ is linear in the spin-orbit coupling
strength \cite{Kavokin01,Gorkov2003}. In general the dephasing
caused by the DM contribution cannot be totally eliminated, but
suggestions have been made for gate pulse shapings that eliminate
it to first order \cite{Burkard02}.

Here we give a different contribution to the spin-spin coupling of
two electrons in III-V semiconductors that does requires neither
bulk inversion asymmetry or Rashba coupling. It arises from the
Coulomb interaction between two electrons and from the
conduction-valence band mixing. It is similar to the exchange
interaction between excitons \cite{BirPikus}. We describe this
asymmetric electron spin coupling in bulk materials, and we make
detailed evaluations for spins in several coupled QD systems.

We use a $8$$\times$$8$ Kane Hamiltonian to represent the band
structure of III-V semiconductors \cite{BirPikus}, with a gap
between an $s$-like conduction band and $p$-like valence bands.
The band coupling in this Hamiltonian is obtained in ${\bf
k}$$\cdot$$\hat{\bf p}$ effective mass perturbation theory, where
$\hat{\bf p}$ is the relevant momentum operator. The standard
parameters in this approach are the band gap $E_g$, the energy of
the split-off band $\Delta$, and the valence-conduction band
coupling $P$=$(\hbar/m_0)\langle S|\hat{p}_x|X\rangle$, where
$|S\rangle$ and $|X\rangle$ are the Bloch states of the conduction
and valence bands \cite{KaneParam}.

We consider two electrons of relative coordinate ${\bf r}$$=$${\bf
r}_1$$-$${\bf r}_2$ in a semiconductor of dielectric permittivity
$\kappa$. The unperturbed two-particle Hamiltonian in the
conduction band is $H^{(0)}$$=$$(\hat{\bf p}_1^2$$+$$\hat{\bf
p}_2^2)/2m_0$. The two band mixing terms ${\bf
k}_1$$\cdot$$\hat{\bf p}_1$, ${\bf k}_2$$\cdot$$\hat{\bf p}_2$,
and  the Coulomb interaction $U_C$$=$$e^2/\kappa {\bf r}$ between
the electrons are treated on equal footing. The leading part of
the spin-dependent interaction arises in third order from first
order contributions of $U_C$ and of each of the band mixing terms
${\bf k}_1$$\cdot$$\hat{\bf p}_1$, ${\bf k}_2$$\cdot$$\hat{\bf
p}_2$. The spin-dependent part of the Coulomb interaction arises
from the coupling between the electron spins and their relative
motion:
\begin{eqnarray}
H^{(3)}_s&=-&\frac{e^2}{\kappa} \frac{2P^2}{3 E_g^2}\frac{\Delta(2
E_g+\Delta)}{(E_g+\Delta)^2} \label{SpinCoulombInt}\\
&&\times\left[({\bf r}\times {\bf p}_1)\cdot {\bf S}_1-({\bf
r}\times {\bf p}_2)\cdot{\bf S}_2\right]/\hbar^2 r^3\,\,.\nonumber
\end{eqnarray}
The coefficient of the square bracket in Eq.(\ref{SpinCoulombInt})
is the coupling constant, which we will call $C$. For GaAs
$C$=$5.7$~meVnm$^3$, and for InAs $C$=$10$~meVnm$^3$. The
remaining spin-independent terms in this order of the theory
contribute to the isotropic exchange $J$. These include a 'local
contact' term of the form
\begin{equation}
H^{(3)}_{c}= -\frac{4\pi e^2 P^2}{3 \kappa}
\frac{(E_g+\Delta)^2+E_g^2}{E_g^2(E_g+\Delta)^2} \delta({\bf
r})\,\,. \label{BulkIntContact}
\end{equation}

Coupling terms similar to $H^{(3)}_s$ and $H^{(3)}_{c}$ are known
for two electrons in the free space \cite{QED} where they arise
from electron-positron band mixing and Coulomb interaction. In the
present case the effect is stronger because the energy gap $E_g$
between electrons and holes in a crystal is much smaller than the
energy gap $m_0$$c^2$ between electrons and positrons. Also, the
symmetry of interaction $H^{(3)}_s$ [Eq.(\ref{SpinCoulombInt})] is
 different from that between electrons in free space \cite{QED}, because the valence bands $\Gamma_8$ and $\Gamma_7$  have a symmetry different from the $s$-symmetry of positrons.

We note here that for itinerant electrons the interaction
$H^{(3)}_s$ [Eq.(\ref{SpinCoulombInt})] flips a spin in an
electron-electron collision, providing a new mechanism for the
relaxation of spin polarization in addition to other known
mechanisms \cite{OpticalPumping}. We evaluated the spin dephasing
time for a 2D electron gas in GaAs quantum wells and found that it
is of the order $\sim$$1$~ns for He temperature and of the order
$\sim$$1$~ps for room temperatures. This suggests that the
interaction $H_s^{(3)}$ can be important in spin transport in
low-dimensional structures, such as quantum wells \cite{Awsch}.

Here we will primarily address systems with spins on two centers,
such as two QDs or two charged donors, where $H^{(3)}_s$
 gives rise to a DM coupling. The two electrons can be in a singlet state with total spin
$S$$=$$0$ or in a triplet state with total spin $S$$=$$1$.
$H^{(3)}_s$ has non-zero matrix elements between states of
different total spin. We take the lowest singlet and triplet
states to be separated by an exchange energy $J$. Then in the
Hilbert space of these two states the Hamiltonian can be written
in the form
\begin{equation}
\tilde{H}=J\,{\bf S}_1\cdot{\bf S}_2+i{\bm \beta}\cdot({\bf
S}_1-{\bf S}_2)\,\,,\label{SpinMixing1}
\end{equation}
where $i{\bm \beta}$=$-C/\hbar^2 \langle S_0|{(\bf r}/
r^3)$$\times$${\bf p}_1|T_0\rangle$ is the matrix element between
the lowest orbital singlet $|S_0\rangle$  and the triplet
$|T_0\rangle$.
A block diagonal form equivalent to
Eq.(\ref{SpinMixing1}) can be obtained by an orthogonal
transformation in spin space, giving the so called 'twisted spin
representation' \cite{Levitov2003}:
\begin{eqnarray}
\tilde{H}&=&J\cos{\phi}\,{\bf S}_1\cdot {\bf S}_2+
2J\left(\sin{\frac{\phi}{2}}\right)^2(\hat{\bf n}\cdot {\bf S}_1)(\hat{\bf n}\cdot {\bf S}_2)+\nonumber\\
&+&J\sin{\phi}\,\hat{\bf n}\cdot({\bf S}_1\times{\bf
S}_2)\label{DM_Hamilt}\,\,,
\end{eqnarray}
where the spin-0 and spin-1 states are mixed by the operator
$\exp{\left[i(\phi/2)\hat{\bf n}\cdot({\bf S}_1-{\bf
S}_2)\right]}$, with ${\bf n}$=${\bm \beta}/\beta$ and
$\phi$$=$$\arctan(\beta/J)$.
In Eq.(\ref{DM_Hamilt}) the first two
terms are the symmetric isotropic and the symmetric anisotropic
Heisenberg terms. The last term is the asymmetric exchange in the
DM form ${\bm \beta}$$\cdot$$({\bf S}_1$$\times$${\bf
S}_2)$$/\sqrt{\beta^2+J^2}$. The contributions from the
Dresselhaus and Rashba couplings to the anisotropic and asymmetric
exchange were given in the form of Eq.(\ref{DM_Hamilt}) in
Ref.\cite{Kavokin01}.

We consider the lowest single-particle states
$|\varphi_\pm\rangle$ from the two confining centers. In the
absence of spin-orbit coupling the two-particle wavefunctions can
be written as products of orbital states and spin states. The
axial vector ${\bm \beta}$ is nonzero only for
inversion-asymmetric confining potentials. For such systems we
take into account the possibility that both electrons occupy the
same site, therefore we use the Hund-Mulliken description of the
two-particle orbital states \cite{Wu02}. We obtain the ground
state by diagonalizing analytically the tunnelling and the Coulomb
and local contact interactions. We then focus on the Hilbert space
determined by the lowest singlet ($S$$=$$0$) and triplet states
($S$$=$$1$) in which the Hamiltonian is given by
Eq.(\ref{SpinMixing1}). The importance of the resulting DM
asymmetric exchange depends on the ratio $\tan{\phi}$=$\beta/J$
between the coefficients of the third and the first terms in
Eq.(\ref{DM_Hamilt}).

\begin{figure}[htbp]
\unitlength1cm
\begin{picture}(8.5,5.7)
 \includegraphics{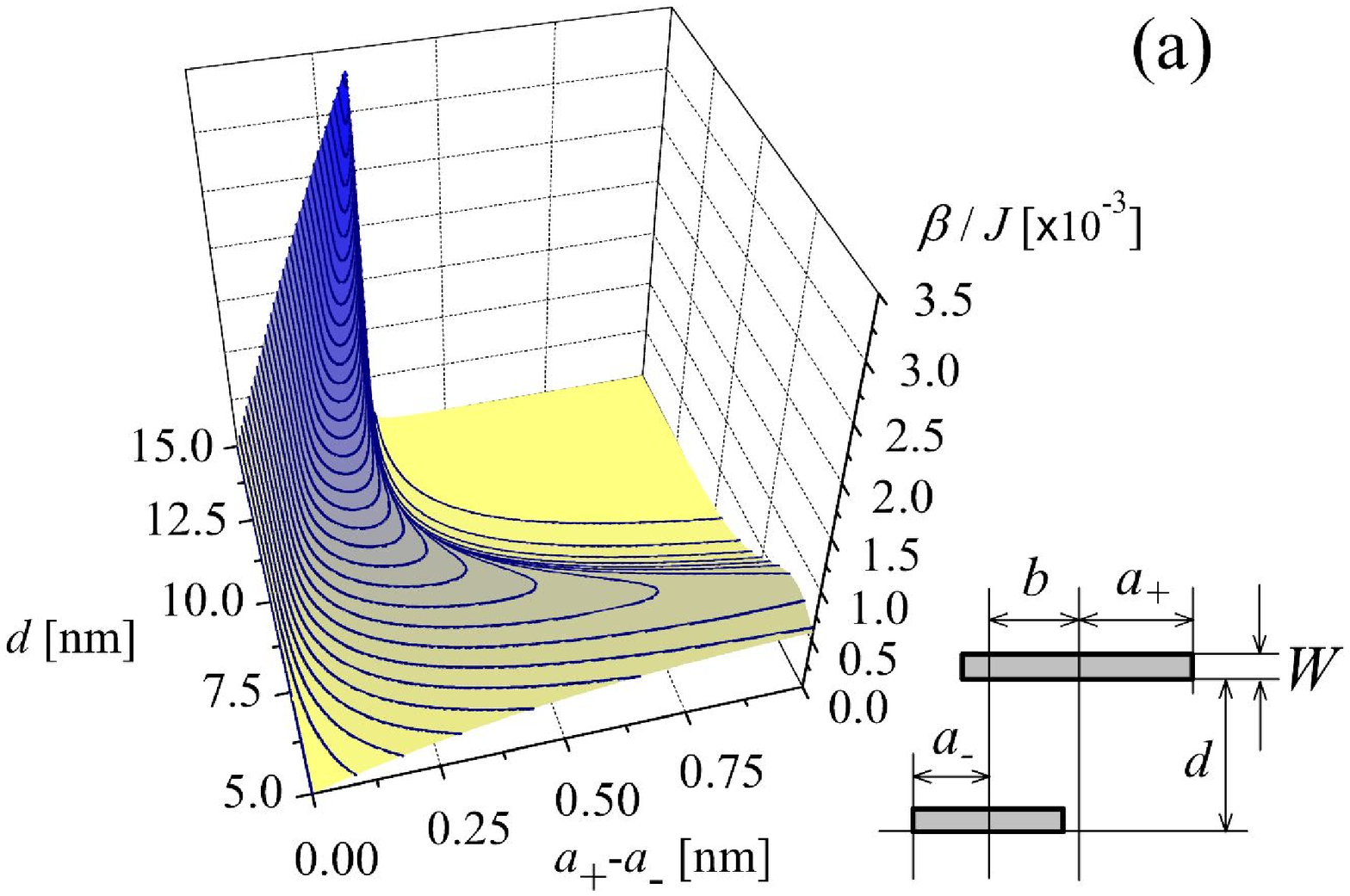}
\end{picture}
\begin{picture}(8.5,5.7)
\includegraphics{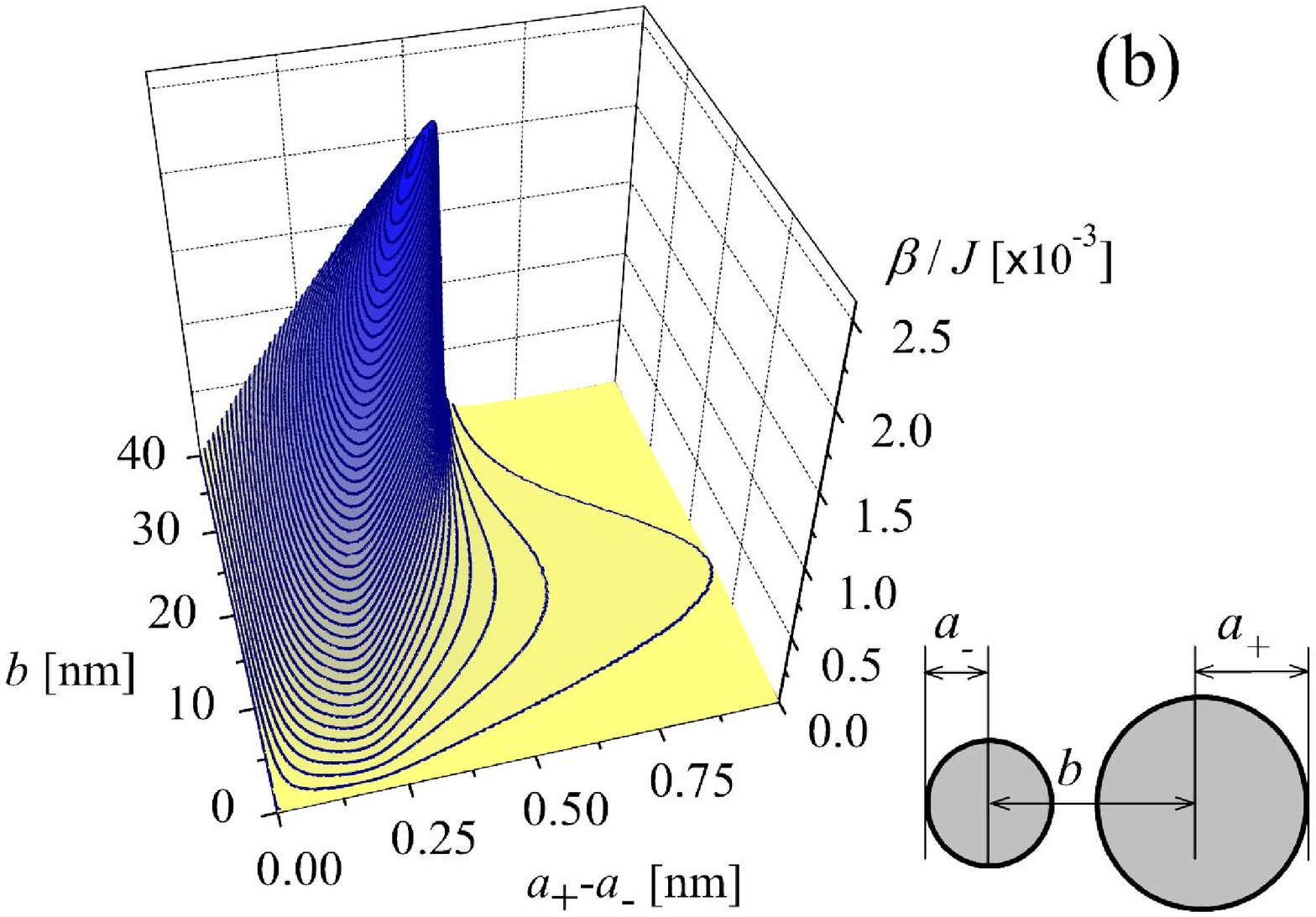}
\end{picture}
\begin{picture}(8.5,4.1)
\includegraphics{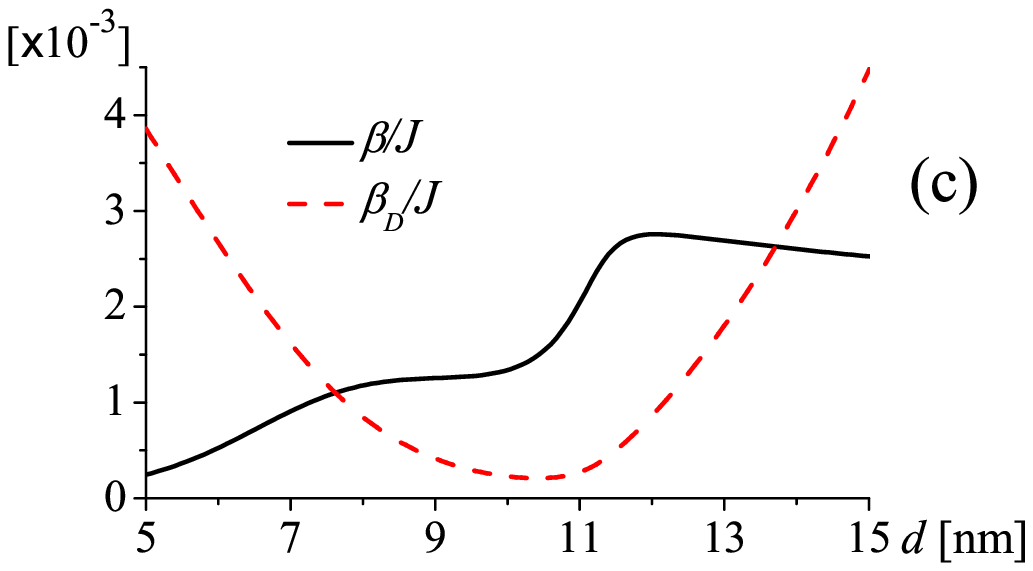}
\end{picture}
\caption{(Color online) (a) Dependence of the asymmetric exchange
$\beta/J$ on the separation $d$ along the growth axis between two
cylindrical vertically-coupled dots with a lateral offset
$b$$=$$4$~nm, for a smaller dot size $a_-$$=$$5$~nm and
differences between radii of dots $\Delta
a$=$a_+$-$a_-$$\in$$\left[5,6\right]$~nm.  (b) dependence of
asymmetric exchange on the lateral offset $b$ at a separation
$d$$=$$10$~nm for $a_-$$=$$5$~nm and varying $\Delta a$. The
insets in (a/b) are lateral/vertical views of the geometry. (c)
Asymmetric exchange $\beta/J$ obtained here (solid line) compared
with the asymmetric exchange $\beta_D/J$ from Dresselhaus coupling
(dashed line) as a function of separation $d$, for $\Delta
a$=$0.22$~nm, $a_-$=$5$~nm, $b$=$4$~nm.} \label{d-dependence}
\end{figure}

Consider first the ratio $\beta/J$ for electrons confined by two
vertically coupled QDs, such as to those in InAs/GaAs systems
\cite{SK}. These dots generally have different sizes and can
 have shape asymmetries as well \cite{AsymDots}. The band offset between GaAs and
InAs ($U_0$$\sim$$0.7$~eV) is larger than the quantization energy
in the lateral direction ($\hbar\omega$$\sim$$20$~meV) by more
than an order of magnitude. This allows us to decouple the
vertical and lateral degrees of freedom. We describe these dots by
the potential offsets along the growth axis $z$ and parabolic
potentials in lateral directions $x$ and $y$, which in general can
be anisotropic. These lateral directions are independent of the
crystal axis. The potentials and wavefunctions then can be written
straightforwardly \cite{VertDots,GaussianModel}. Thus, the
asymmetry studied here results from
 differences in the lateral sizes and shapes of the QDs.

The dependence of the ratio $\beta/J$ of the asymmetric exchange
to the symmetric exchange on the separation $d$ between the two
cylindrical dots with a fixed lateral offset $b$ is shown in
Fig.\ref{d-dependence}(a). Its dependence on the lateral offset
$b$ for two dots at a fixed $d$ is shown in
Fig.\ref{d-dependence}(b). The insets of
Fig.\ref{d-dependence}(a,b) show sketches of the lateral and
vertical views of two cylindrical dots with vectors ${\bm b}$,
${\bm d}$ pointing from the smaller dot to the larger one. The
vector ${\bm \beta}$ is oriented in the direction ${\bm
b}$$\times$${\bm d}$, and for small differences $\Delta
a$=$a_+$$-$$a_-$ between the dot radii $a_\pm$, it is proportional
to $\Delta a$.  It follows from Eq.(\ref{SpinCoulombInt}) that for
two cylindrical dots of equal sizes $\beta/J$ vanishes because the
system then has a center of inversion and the ground state is
symmetric is no asymmetric exchange because the system has
cylindrical symmetry (${\bm b}$$\times$${\bm d}$=0). In each of
these figures there are two regions of behavior: (i) For small
size differences $\Delta a$ (the left sides of the peaks)
$\beta/J$ increases roughly proportional to $d$ in
Fig.\ref{d-dependence}(a) and roughly proportional to $b$ in
Fig.\ref{d-dependence}(b) until it reaches
$\sim$$3.5$$\times$$10^{-3}$; in this region the two electrons are
distributed almost symmetrically between the dots. (ii) For larger
values of $\Delta a$ (right sides of the peaks), $\beta/J$ first
increases with $d$ to a small maximum and then decreases
[Fig.\ref{d-dependence}(a)], and similarly with $b$
Fig.\ref{d-dependence}(b)]. In this second region both electrons
tend to occupy preferentially the larger dot, where the Coulomb
energy is overcome by the difference between single-particle
energies of the two dots.

\begin{figure}[htbp]
\unitlength1cm
\begin{picture}(8.5,5.5)
\includegraphics{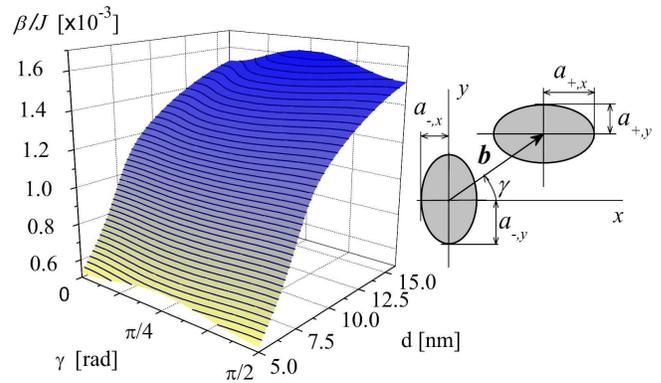}
\end{picture}
\caption{(Color online) Dependence of modulus of asymmetric
exchange $\beta/J$ on angle $\gamma$ between the axes of two
elliptical vertically-coupled dots of equal major and minor axes
$a_{-,x}$$=$$a_{+,y}$$=$$3.5$~nm,
$a_{-,y}$$=$$a_{+,x}$$=$$6.5$~nm, at $b$=$4$~nm. The inset gives a
projection of the dot contours on a plane perpendicular to the
growth axis, using the harmonic potential parameters to represent
the dot sizes.} \label{Phi-dependence2}
\end{figure}

In order to study shape asymmetry, we consider two vertically
coupled dots with deviations from cylindrical symmetry. In
Fig.\ref{Phi-dependence2} we give results for two identical
elliptical dots that are rotated by $\pi/2$ with respect to one
another, as shown in the inset, with an offset $b$=$4$~nm. The
dots of equal sizes have equal energies, which leads to an equal
distribution of two electrons on them. The ratio $\beta/J$ reaches
substantial values ($\sim$$10^{-3}$) as a function of the
separation $d$ and has a relatively weak dependence on the angle
$\gamma$ between the axis of the dots. In this case $\beta$ arises
from the shape asymmetry. The coupling depends on the angle
$\gamma$ between the relative position vector and the principal
axis (for cylindrical dots, ${\bm \beta}$ changes direction but is
constant in magnitude). The orientation of ${\bm \beta}$ is given
by ${\bm b}$$\times$${\bm d}$. In the case where the axies of the
dots are parallel ($a_{-,x}$=$a_{+,x}$, $a_{-,y}$=$a_{+,y}$), the
asymmetric exchange is zero (the system then has an inversion
center at $({\bm b}/2,{\bm d}/2)$). In cases when the dots are
different in size and not cylindrical,
 $\beta/J$ is even larger.
For example, for a cylindrical dot with $a_-$=$5$~nm coupled with
an elliptical dot with $a_{+,x}$=$4$~nm, $a_{+,y}$=$6.25$~nm, the
maxima of $\beta/J$ are $\sim$$3.5$$\times$$10^{-3}$ for all
$\gamma$.

\begin{figure}[htbp]
\unitlength1cm
\begin{picture}(8.5,5.6)
\includegraphics{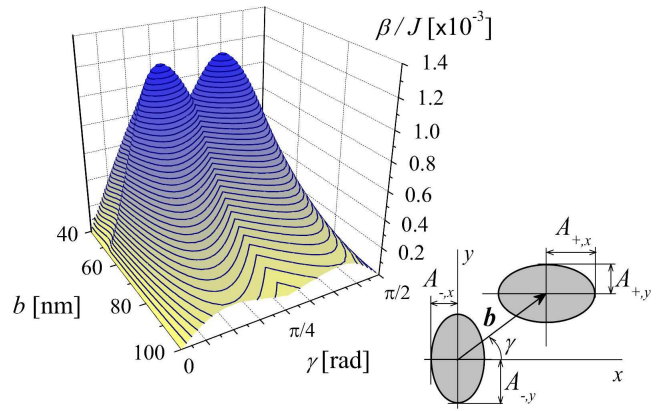}
\end{picture}
\caption{(Color online) Dependence of modulus of asymmetric
exchange $\beta/J$ on angle $\gamma$ between the axes of two
elliptical laterally-coupled dots of equal potential sizes
$A_{-,x}$=$A_{+,y}$=$6$~nm and $A_{-,y}$=$A_{+,x}$=$9$~nm. The
inset shows the two QDs that lie in the same plane, seen along the
growth axis, using the Gaussian potential parameters to define
their sizes. } \label{Phi_dependence_lateral}
\end{figure}

We have calculated the contribution to the asymmetric exchange for
these structures from the bulk Dresselhaus coupling
$H_{so}^D$$=$$i\gamma_{so}^{D}\partial_x(\partial_y^2-\partial_z^2)$$S_\alpha/\hbar$
(plus cyclic permutations of cartesian indices). The coupling
constant $\gamma_{so}^{D}$ is $47$~meVnm$^3$ for GaAs, and of the
order of $100$~meVnm$^3$ for InAs. In Fig.\ref{d-dependence}(c) we
compare the contribution $\beta_D/J$ to the asymmetric exchange
from the Dresselhaus coupling with asymmetric exchange $\beta/J$
for a size difference $\Delta a$=$0.22$~nm \cite{Compare_AE_D,
linearDresselhaus,Rashbaparam_InAs}. We see that $\beta/J$ is
larger for intermediate separations $d$, a region of particular
interest for implementations for quantum information. As the
difference in the dot sizes increases, $\beta/J$ becomes larger
relative to $\beta_D/J$, and as the size difference decreases,
$\beta_D/J$ becomes larger.

We have also considered laterally coupled QDs. To obtain a
representation of the barrier between the QDs we use
inverted-gaussian potentials \cite{ReviewQD02}, and we use again
the material parameters for InAs. The wavefunctions are obtained
variationally \cite{LatDots}. In
Fig.(\ref{Phi_dependence_lateral}) we give results for two
elliptical dots of equal sizes rotated with respect to each other
by $\pi/2$. In this case the anisotropic exchange $\beta/J$ arises
from the shape asymmetry. From the operator ${\bf r}$$\times$${\bf
\nabla}_1$ in Eq.\ref{SpinCoulombInt}, the asymmetric exchange
${\bf \beta}$ has a nonzero component only along the growth axis,
and the dependence of the modulus $\beta/J$ is symmetric with
respect to $\gamma$$=$$\pi/4$. In
Fig.(\ref{Phi_dependence_lateral}) once again $\beta/J$ reaches a
maximum of $\sim$$10^{-3}$. In cases when the Dresselhaus
\cite{linearDresselhaus} and Rashba \cite{Rashbaparam_InAs}
couplings have equal coupling constants, their contribution is
small for $\gamma$=$\pi/4$ and then the total asymmetric exchange
is dominated by $\beta/J$.

In summary, we derived an asymmetric contribution to the exchange
interaction between two electrons in III-V semiconductors that
arises from the Coulomb interaction and the band mixing and does
not require inversion asymmetry. For asymmetric coupled
semiconductor QDs, this contribution depends on the geometry and
is typically
 $10^{-3}$ of the isotropic exchange $J$.  This interaction also can play a role in the relaxation and dephasing of spin in transport processes in low-dimensional structures.

This work was supported by the DARPA QUIST program, by ONR and by
the ONR Nanoscale Electronics Program.


\begin{thebibliography}{200}
\bibitem{KaneParam}The Kane parameters used here are: $\Delta$$=$$0.39$~eV, $E_g$=$0.42$~eV, $2P^2/m_0$=$21.5$~eV for InAs and $\Delta$=$0.417$~eV,
$E_g$=$0.39$~eV, $2P^2/m_0$=$28.8$~eV for GaAs, where $m_0$ is the
free electron mass. The conduction band of III-V semiconductors
has symmetry $\Gamma_6$ and the valence bands have symmetries
$\Gamma_7$ and $\Gamma_8$.
\bibitem{VertDots}The QDs are centered at $\pm(b_{x}/2,b_{y}/2,d/2)$. Using $x_\pm$=$\frac{x\mp b_x}{2}$, $y_\pm$=$\frac{y\mp b_y}{2}$,
$z_\pm$=$\frac{z\mp d}{2}$ for the two dots ($+/-$), the
single-dot potentials  are (Ref.\cite{GaussianModel})
$U_{\pm}^v$=$\left[m (\omega^2_{\pm,x}x_{\pm}^{2}+
\omega^2_{\pm,y}y_{\pm}^{2})/2-U_0\right]$$W\delta(z_\pm)$. For
each separate dot, the lowest one-particle electronic state is
$\varphi_{\pm}^{0}$=$\sqrt{\frac{q}{4\pi a_{\pm,x}
a_{\pm,y}}}$$\exp\left[- \left(\frac{x_\pm}{2
a_{\pm,x}}\right)^2-\left(\frac{y_\pm}{2
a_{\pm,y}}\right)^2-\frac{q|z_{\pm}|}{2}\right]$. The lateral
sizes of $\varphi_{\pm}^{0}$ are
$a_{\pm,\alpha}$=$\sqrt{\frac{\hbar}{2m\omega_{\pm,\alpha}}}$ and
the vertical size is $q$=$\frac{2 m_b W U_0}{\hbar^2}$, where:
$\alpha$=$x,y$, $\hbar \omega_{\pm,\alpha}$ are the lateral
quantization energies and $m$=$0.023$~$m_0$, $m_b$=$0.067$~$m_0$
are the electron effective masses in InAs and in the GaAs barrier.
The total confining potential is $U_t^v$=$U_+^v+U_-^v$. The two
Wannier states $|\varphi_{\pm}\rangle$ are obtained by a symmetric
orthogonalization of the single-dot orbitals:
$|\varphi_{\pm}\rangle$$=$$(\cos
{\frac{\theta}{2}}|\varphi_{\pm}^{0}\rangle-\sin
{\frac{\theta}{2}} |\varphi_{\mp}^{0}\rangle)/\cos{\theta}$. We
consider dots of equal thicknesses $W$, and evaluate the lateral
potentials from splittings between the lowest states in
experiment.
\bibitem{Compare_AE_D} In general, a structure asymmetry can give
rise to a Rashba coupling that can be opposite in sign to the
Dresselhaus coupling \cite{linearDresselhaus} and thus decrease
$\beta_D/J$. This contribution is expected to be small for the
present system.
\bibitem{LatDots}The inverted Gaussian potentials $U_\pm^l$ of widths $A_{\pm,\alpha}$ for the two QDs ($+/-$)
and the  wavefunctions are, respectively:
$U_\pm^l$=$V_\pm\exp\left[-\left(x_\pm/A_{\pm,x}\right)^2-\left(y_\pm
/A_{\pm,y}\right)^2\right]$,
$\varphi_{\pm}^{0}$=$\exp\left[-\left(x_\pm/2
a_{\pm,x}\right)^2-\left(y_\pm/2
a_{\pm,y}\right)^2\right]/\sqrt{2\pi a_{\pm,x} a_{\pm,y} }$, where
$A_{\pm,\alpha}$ define the size of the dots and the wavefunction
parameters $a_{\pm,\alpha}$ are obtained variationaly.
\end{thebibliography}
\end{document}